# 7-Rod-Core Thulium-Doped Fiber for Enhanced Fiber Laser Cooling

**Ivo Bartoň**[1*], **Jan Aubrecht**[1], **Bara Švejkarová**[1,2], **Jan Pokorný**[1,2], **Michal Kamrádek**[1], **Ondřej Podrazký**[1], **Petr Vařák**[1], **Ivan Kašík**[1], **Martin Grábner**[1], **Pavel Peterka**[1]

[1)]Institute of Photonics and Electronics of the Czech Academy of Sciences, Chaberská 1014/57, 182 00 Prague, Czechia
[2)]Faculty of Nuclear Sciences and Physical Engineering, Czech Technical University in Prague, Břehová 7, 115 19 Prague, Czechia
*barton@ufe.cz

Abstract: Structured-core thulium-doped fibers were developed to reduce heat load, enable shorter-wavelength operation, and achieve a pedestal-free design. In a proof-of-principle experiment, laser slope efficiencies of 52% at 1907 nm and 54% at 1940 nm were achieved with respect to absorbed power.

## 1. Introduction

Thulium-doped fiber lasers (TDFLs) and amplifiers emitting near 2 μm in the eye-safe infrared region are widely used in medicine, sensing, and frequency conversion because of their low risk of retinal damage and strong water absorption [1]. They can be efficiently pumped at 0.79 μm using commercial diodes that excite the $^3H_4$ energy level. Although the Stokes limit restricts the theoretical efficiency to about 40%, TDFLs benefit from a cross-relaxation "2-for-1" process ($^3H_4$, $^3H_6$ → $^3F_4$, $^3F_4$) enabled by high thulium ion ($Tm^{3+}$) concentrations (>7000 ppm), which significantly enhances laser efficiency [2]. Tm-doped fibers (TDFs) for high-power lasers are often made using modified chemical vapor deposition (MCVD) combined with solution doping methods to incorporate both $Al_2O_3$ and $Tm^{3+}$ into the fiber core [3]. For high-power lasers with large mode areas, high doping concentrations require the so-called pedestal design to lower the effective numerical aperture (NA) and maintain single-mode operation. Since the pedestal has a lower effective NA in the fiber core, it acts as a high-NA waveguide itself, serving as both cladding for the active core and as a secondary, redundant core [4]. This secondary core is formed because the pedestal region is surrounded by a lower-refractive-index pure silica layer that serves as the standard fiber cladding. Consequently, the presence of a pedestal can complicate fiber splicing and reduce fiber laser reliability. A fraction of the light that cannot be guided in the active core due to external perturbations or coupling issues becomes trapped in the pedestal and may negatively impact laser efficiency and reliability [5]. Recent research has focused on improving fabrication methods to maximize doping concentrations, such as vapor-phase deposition [6] or nanoparticle-doping techniques [7], as well as optimizing fiber design for power scaling and reliability [8], including pedestal-free designs [9,10]. To achieve such a design, the stack-and-draw (SAD) technique—adapted for rare-earth-ion (RE)-doped fibers [11] — enables precise doping of high-concentration areas, facilitating a 2-for-1 process while maintaining a low average concentration across the fiber core. Reducing the average $Tm^{3+}$ concentration also helps prevent overheating, thereby improving reliability [5]. The pedestal-free design presented in [9] has employed SAD to create nanoscale structures within the fiber core. The key advantage of nanostructuring is the ability to independently control the local $Tm^{3+}$ concentration and the fiber core's effective refractive index. Instead of a uniform distribution, $Tm^{3+}$ would be concentrated into nanorods. It also enables control over the core's effective refractive index by adjusting nanorod spacing when filling passive regions with pure silica. Optimizing the size, doping level, and spacing of nanorods permits control over the refractive index difference between the core and cladding without using any pedestal area. In the case of a nanostructured fiber [9], a relatively low slope efficiency of 29% was achieved, partially due to the low concentration of $Tm^{3+}$ in the nanorods, which was close to the threshold for cross-relaxation and due to the dopant diffusion. Additionally, the combination of a low average $Tm^{3+}$ concentration across the fiber cross-section and high background losses at a level of 200 dB/km prevented higher efficiency.

In this work, we present a proof-of-concept for fabricating novel TDFs with a structured core and a high local concentration of $Tm^{3+}$ (13000 mol ppm). Using the SAD technique, two fibers were fabricated: the first featuring a core of seven $Tm^{3+}$-doped elements, and the second with a ring structure of $Tm^{3+}$-doped elements surrounding an $Al_2O_3$ element. Comparing the optical and laser properties of both fiber lasers shows that they can achieve 52% SE at 1907 nm relative to absorbed power and 54% SE at 1940 nm.

## 2. Fiber preparation

Two types of initial preforms were prepared similarly to [7,12] using the MCVD method combined with nanoparticle doping: one doped with $Tm^{3+}$ and $Al_2O_3$, and the other with only $Al_2O_3$. The preforms were etched in concentrated hydrofluoric acid to remove the silica cladding and then elongated into 330 μm canes. F300 tubes and rods were drawn to appropriate diameters, and using the SAD technique, the final preform was assembled according to the designs illustrated in Fig. 1A and 1B. The fiber core was assembled using 330 μm canes doped with $Tm^{3+}$ (blue circles) and $Al_2O_3$ (red circle), as shown in the insets of Fig. 1A and 1B. The assembled final preforms were drawn into fibers at 1900 °C. The fibers were designated TDF-7Tm and TDF-6x1Ring.

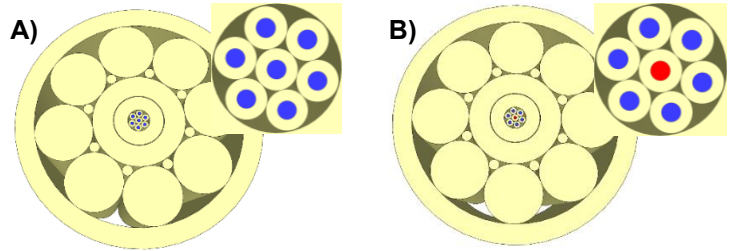

Fig. 1. A) Design of TDF-7Tm, B) Design of TDF-6x1Ring.

## 3. Characterization of fibers and laser measurements

The overall macroscopic appearance of the fiber end face and detailed characterization of the fiber's core were obtained using the Scanning electron microscope (SEM) device TESCAN LYRA3 GM with a backscattered electron detector; the fiber sample was coated with a 20-nm-thick layer of carbon. The appearance of the face of the drawn TDFs can be seen in Fig. 2. The fiber's core has a diameter of approximately 10 μm, with a silica cladding measuring 131 μm. In Fig. 2A, the core contains seven separated bright spots representing $Tm^{3+}$-doped regions. The core diameter was measured as the diameter of a circumscribed circle around all seven rods. A closer look at the fiber core through SEM is shown in the inset of Fig. 2A, which indicates that the $Tm^{3+}$-doped areas have a diameter of about 1.91 μm. Figure 2B displays six bright spots surrounding one darker area of $Al_2O_3$. This contrast results from the presence of heavier $Tm^{3+}$ ions compared to the undoped $Al_2O_3$ region. The inset in Fig. 2B, showing the fiber core, clearly reveals a ring structure.

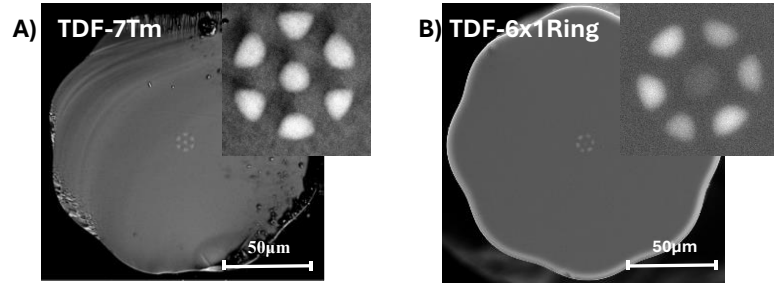

Fig 2. SEM of whole fiber with an inset of a close look at the core of fiber A) TDF-7Tm, B) TDF-6x1Ring.

The fibers' refractive index profile (RIP) was measured using an optical fiber analyzer IFA-100 (Interfiber Analysis Inc.) with a laser source emitting at 979 nm

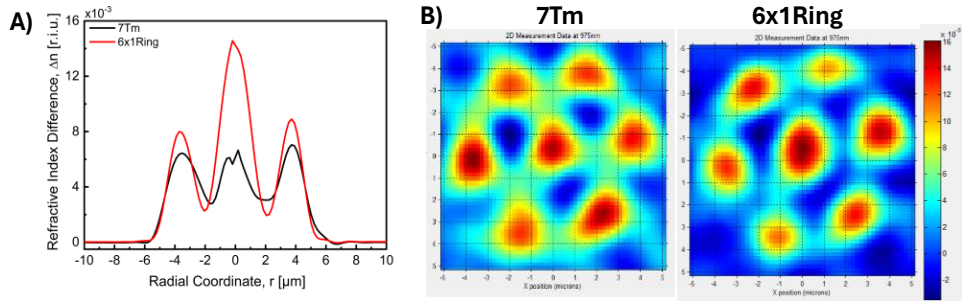

Fig 3. A) 1D RIP of TDF-7Tm and TDF-6x1Ring, B) 2D RIP TDF-7TM and TDF-6x1Ring.

A 1D scan and 2D tomography were performed. For TDF-7Tm, the refractive index difference (RID) between the doped areas and the cladding is 7×10-3, with a core diameter of approximately 9.5 μm. The 2D tomography zoomed into the core area displays seven distinct sections representing Tm-doped areas. For TDF-6x1Ring 1D (see Fig. 2A), the average RID is 8×10-3, with the central area reaching 14.5×10-3 (Fig. 2A). This matches the 2D tomography, which shows a ring structure (Fig. 2B). Note that the IFA spatial resolution is about 0.5 μm, which clearly influences the resolution of individual rods in the core.

Both fiber cores were characterized by their spectral absorption. A tungsten-halogen lamp was used as a broadband radiation source. To determine the core absorption, we employed a modified cutback method [13], where the transmission properties of the corresponding section of the active fiber, which was pigtailed from both sides with suitable passive fibers, and a reference fiber of the same type as the pigtails, were measured and then compared. Absorption spectra were recorded with a Thorlabs OSA203C spectrometer over 1000–2600 nm, with a resolution of approximately 1 rcm or 0.4 nm. The measured absorption spectra of both TDFs are shown in Fig. 3A. For both fibers, absorption can be described as:

$$\text{Abs}(\lambda) = 4.34 \times (\sigma a^{Tm} \Gamma(\lambda) N_{total}^{Tm}) \qquad (1)$$

where $\Gamma$ corresponds to the overlap factor, $\sigma a^{Tm}$ is the absorption cross-section, and $N_{total}^{Tm}$ stands for the total concentration of $Tm^{3+}$ in the fiber core. The shape of peaks depicted in Fig. 4A is comparable with absorption peaks stated in the literature for Thulium-doped fiber lasers [14,15], confirming the presence of $Tm^{3+}$ in the fiber core. Ground-state absorptions were measured at 99.3 dB/m at 1640 nm for TDF-7Tm and 85.56 dB/m for the TDF 6+1 Ring structure. These values are comparable to those obtained for MCVD solid-core thulium fiber lasers used as broadband amplifiers [14].

Background losses (attenuation) were measured using the standard cut-back method with a long fiber piece of 184 m for TDF-7Tm and 132 m for TDF-6+1 and a short fiber piece of 2 m for both fibers under the constant launched condition. A broadband halogen lamp, bare fiber adapters, and a monochromator-based spectrometer (Ando, AQ6317B with a spectral resolution of 2 nm) were used. Excitation of the whole face of the fiber was used during the characterization. Measured minimum values of attenuation of 66 dB/km for TDF-7Tm and 55 dB/km for TDF-6x1Ring, both at 840 nm, show that these values are comparable with fibers produced by MCVD [16] and approximately 3.5 times lower compared to nanostructured fiber [9].

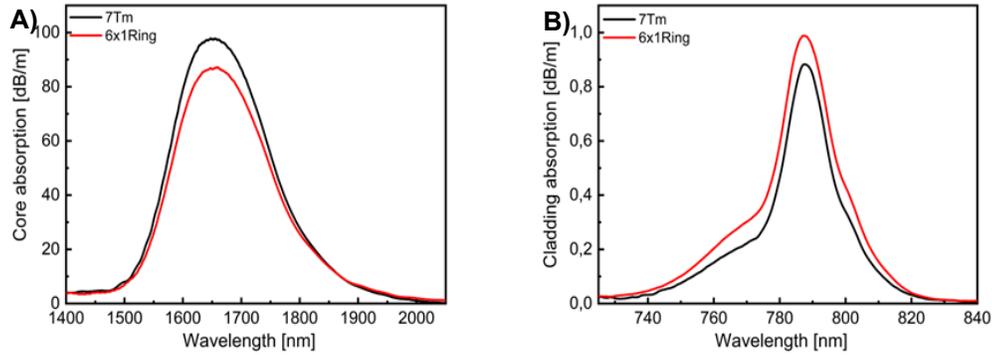

Fig 4. A) Core absorption of TDF-7Tm and TDF-6x1Ring, B) Cladding absorption of TDF-7TM and TDF-6x1Ring.

Cladding absorption is shown in Fig. 4B. The spectral region around the emission wavelength of the laser diode (around 790 nm) was measured using a broadband tungsten-halogen lamp, and transmitted intensities were detected with an Ando AQ6317B monochromator-based spectrometer over 600-900 nm, with a spectral resolution of 2 nm.

For cladding absorption measurements, several-meter-long sections were spliced to passive multimode fibers to excite the larger cladding regions, and a modified cutback method was used. Subsequently, the cladding absorption values can be used to estimate the optimal length of the laser cavity. Relatively low absorption values (<1 dB/m) indicated the need to use longer resonators (~> 10 m).

The samples were characterized by fluorescence lifetime according to the procedure shown in detail in [16]. The $^3F_4$ lifetime, corresponding to emission near 2 μm, was measured under in-band pumping at 1620 nm using a diode source. The emission was detected by an InGaAs (Hamamatsu) photodiode. The measured fibers were approximately 1 mm long, and emission was detected from the side in order to suppress the influence of amplified spontaneous emission and reabsorption.

Measured lifetimes for different pumping powers are shown in Fig. 5A and 5C for TDF 7Tm and TDF 6+1 Ring, respectively. Decay times at different pumping powers are depicted in Fig. 5B and 5D for TDF 7Tm and TDF 6+1 Ring, respectively. From the obtained measurements, it can be seen that at low pump powers, the fluorescence decay exhibited single-exponential behavior, indicating a homogeneous distribution of $Tm^{3+}$ ions within both fiber cores. As the excitation power increased, the decay curves deviated from this behavior, and the measured decay time, obtained from the 1/e intensity on the normalized curve, shortened, which can be attributed to the growing influence of energy-transfer (ET) processes between neighboring ions. The measured decay time saturated at low powers, and the fluorescence lifetime extrapolated to zero power was nearly identical in both samples, ranging from 515 to 520 μs. As stated in [16], fluorescence lifetimes above 500 μs in alumino-silicate fibers generally indicate a favorable low-phonon environment for Tm³⁺ ions and show promise for efficient laser operation.

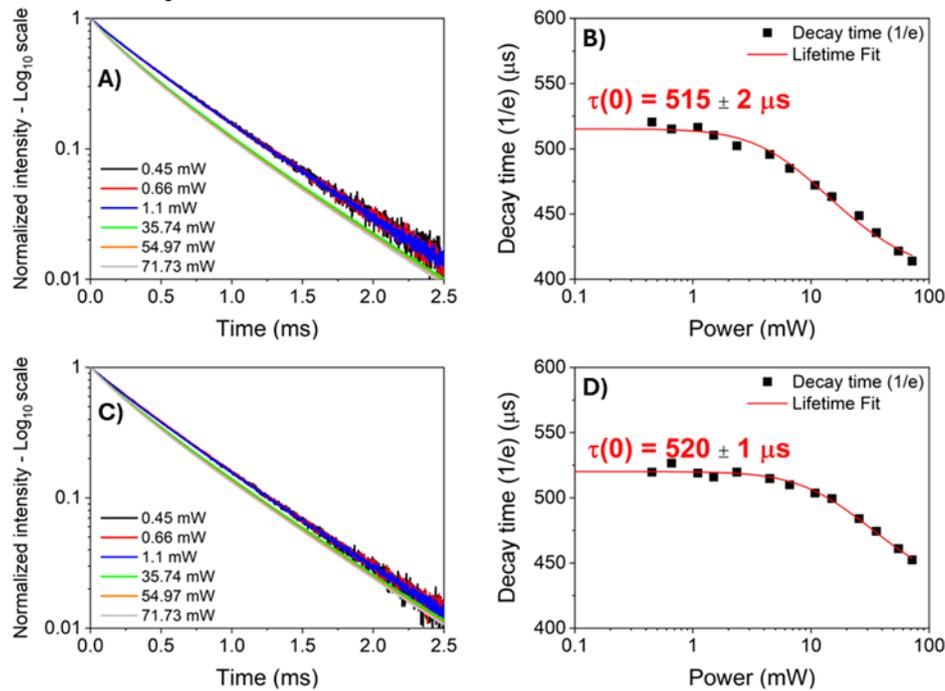

Fig 5 A)Lifetimes of TDF-7TM at different pumping powers B) Fluorescence decay times at different pumping powers of TDF-7Tm , C) Lifetime of TDF-6x1Ring at different pumping powers D)Fluorescence decay times at different pumping powers of TDF-6x1Ring. The decay times were fitted for both fibers using Equation 1 in reference [15].

Laser performance of structured-core fibers was tested in a Fabry-Perot laser configuration, as depicted in Fig. 6. The TDFs were pumped with a 792 nm laser diode (BWT, 105/125 μm), with limited pump power up to 45 W. A high-reflectivity fiber Bragg gratings were used, reflecting at 1907 nm (Advanced Fiber Resources, 10/130 μm) and at 1940 nm (TeraXion, 10/130 μm), respectively. The output coupler was created by Fresnel reflection (~ 4 %) from a perpendicularly cleaved fiber end. The output power was measured using a dichroic mirror (Crytur), Thorlabs S425C-L, and a Thorlabs S142C power meter.

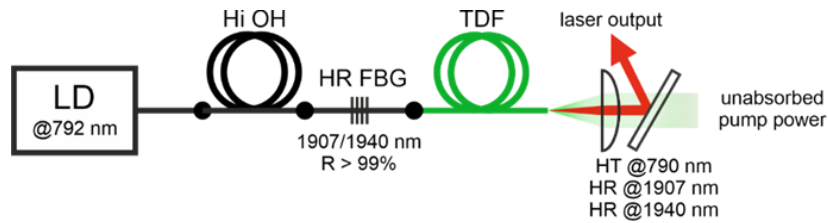

Fig 6. Experimental set-up for measurement of laser performance of TDF-7Tm and TDF-6x1Ring,

Fig. 7A shows the laser output power for TDF-7Tm as a function of absorbed pump power at 1907 and 1940 nm. At 1940 nm, the maximum output power reached 24.8 W, with a slope efficiency of 54% and a laser threshold of 1.8 W. At 1907 nm, the maximum output power reached 15.6 W, with an SE of 51.9% and the same laser threshold of 1.8 W. Very similar values were obtained for TDF-6x1ring, as shown in Fig. 7B. At 1940 nm, the maximum output power reached 24.7 W, with an SE of 54.8% and a laser threshold of 1.8 W. At 1907 nm, a maximum output power of 16.58 W was achieved at the same pump power, with a laser threshold of 1.70 W and a slightly lower SE of 51.8%. No visible roll-off, saturation, or parasitic lasing was observed in any of the laser curves or laser spectra, indicating that the maximum output power can be further increased with a more powerful pump source.

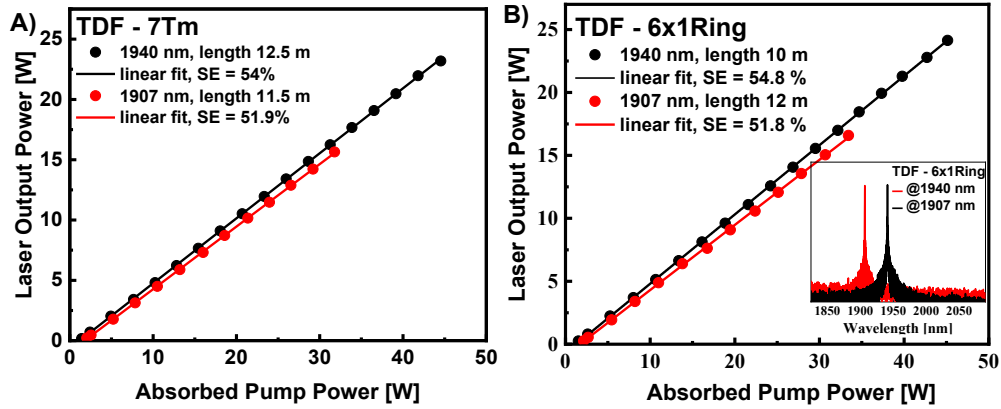

Fig. 7. A) Laser performance with respect to absorbed power for TDF 7Tm B) Laser performance with respect to absorbed power for TDF-6x1Ring

**Table. 1. Values of slope efficiencies for both TDFs at optimal length for 1907nm and 1940nm**

|        | F-P 1907 nm     | F-P 1940 nm    |
|--------|-----------------|----------------|
| 7Tm    | 51.9% / 11.5 m  | 54% / 12.5 m   |
| 6+1Ring| 51.8% / 12 m    | 54.8% / 10 m   |

The optimized fiber lengths were similar, ranging from 10 to 12 m, as shown in Table 1, along with SE values for each fiber and wavelength. These fiber lengths were determined by the relatively low cladding pump absorption of about 1.1 dB/m at 793 nm. Despite a longer resonator, no parasitic lasing was observed, and laser emission remained stable. These SE values are higher than the 29% SE obtained for a nanostructured thulium-doped fiber [9]. Together with lower background losses and a reduced effect of RE diffusion when using much larger $Tm^{3+}$ doped rod elements, these values demonstrate that the proposed structure exhibits promising properties suitable for high-power applications, even at shorter wavelengths

## 4. Conclusions

In this work, two TDF designs were successfully fabricated using the MCVD method with nanoparticle doping and the stack-and-draw technique. Both fiber types exhibited well-defined pedestal-free refractive index profiles. Laser experiments using a Fabry–Perot configuration showed stable emission at 1907 nm and 1940 nm, with SE reaching 52% at both wavelengths. The absence of parasitic lasing, even with extended resonator lengths, confirms the effectiveness of the design. These results demonstrate the potential of pedestal-free, structured cores to lower heat load and improve cooling and reliability of the laser while keeping high fiber laser efficiency, thus offering a promising fabrication route for high-power fiber lasers in the 2 μm spectral region.

**Funding:** Grantová Agentura České Republiky (26-21543S): European Union and the state budget of the Czech Republic(LasAPP CZ.02.01.01/00/22_008/0004573)

**Acknowledgement:** This work was supported by the Czech Science Foundation (26-21543S) and co-funded by the European Union and the state budget of the Czech Republic under the project LasApp CZ.02.01.01/00/22_008/0004573.

**Disclosures:** The authors declare no conflicts of interest

**Data Availability Statement.** Data underlying the results presented in this paper are available online at DOI: (to be filled upon acceptance)

## References

1. M. N. Zervas and C. A. Codemard, "High Power Fiber Lasers: A Review," IEEE Journal of Selected Topics in Quantum Electronics **20**(5), 219–241 (2014).
2. M. Grábner, B. Švejkarová, J. Aubrecht, and P. Peterka, "Analytical Model of Thulium-Doped Fiber Laser Pumped by Two-for-One Process," J. Lightwave Technol. **42**(8), 2938–2944 (2024).
3. A. Sincore, J. D. Bradford, J. Cook, L. Shah, and M. C. Richardson, "High Average Power Thulium-Doped Silica Fiber Lasers: Review of Systems and Concepts," IEEE J. Select. Topics Quantum Electron. **24**(3), 1–8 (2018).
4. K. Tankala, B. Samson, A. Carter, J. Farroni, D. Machewirth, N. Jacobson, U. Manyam, A. Sanchez, M.-Y. Chen, A. Galvanauskas, W. Torruellas, and Y. Chen, "New developments in high power eye-safe LMA fibers," in A. J. W. Brown, J. Nilsson, D. J. Harter, and A. Tünnermann, eds. (2006), p. 610206.
5. M. Michalska, P. Honzatko, P. Grzes, M. Kamradek, O. Podrazky, I. Kasik, and J. Swiderski, "Thulium-Doped 1940- and 2034-nm Fiber Amplifiers: Towards Highly Efficient, High-Power All-Fiber Laser Systems," J. Lightwave Technol. **42**(1), 339–346 (2024).
6. N. J. Ramírez-Martínez, M. Núñez-Velázquez, A. A. Umnikov, and J. K. Sahu, "Highly efficient thulium-doped high-power laser fibers fabricated by MCVD," Opt. Express **27**(1), 196 (2019).
7. M. Kamrádek, I. Kašík, J. Aubrecht, J. Mrázek, O. Podrazký, J. Cajzl, P. Vařák, V. Kubeček, P. Peterka, and P. Honzátko, "Nanoparticle and Solution Doping for Efficient Holmium Fiber Lasers," IEEE Photonics Journal **11**(5), 1–10 (2019).
8. I. Barton, M. Franczyk, P. Peterka, J. Aubrecht, P. Vařák, M. Kamrádek, O. Podrazký, R. Kasztelanic, R. Buczynski, and I. Kašík, "Optimization of erbium and ytterbium concentration in nanostructured core fiber for dual-wavelength fiber lasers," in *Specialty Optical Fibres*, K. Kalli, A. Mendez, and P. Peterka, eds. (SPIE, 2023), **12573**, p. 1257311.
9. P. Peterka, J. Aubrecht, D. Pysz, M. Franczyk, O. Schreiber, M. Kamrádek, I. Kasik, and R. Buczyński, "Development of pedestal-free large mode area fibers with $Tm^{3+}$ doped silica nanostructured core," Opt. Express **31**(26), 43004 (2023).
10. M. J. Barber, P. C. Shardlow, P. Barua, J. K. Sahu, and W. A. Clarkson, "Nested-ring doping for highly efficient 1907 nm short-wavelength cladding-pumped thulium fiber lasers," Opt. Lett. **45**(19), 5542 (2020).
11. W. J. Wadsworth, R. M. Percival, G. Bouwmans, J. C. Knight, and P. St. J. Russell, "High power air-clad photonic crystal fibre laser," Opt. Express **11**(1), 48–53 (2003).
12. M. Kamrádek, J. Aubrecht, P. Vařák, J. Cajzl, V. Kubeček, P. Honzátko, I. Kašík, and P. Peterka, "Energy transfer coefficients in thulium-doped silica fibers," Opt. Mater. Express **11**(6), 1805–1814 (2021).
13. B. Jiříčková, M. Grábner, C. Jauregui, J. Aubrecht, O. Schreiber, and P. Peterka, "Temperature-dependent cross section spectra for thulium-doped fiber lasers," Opt. Lett. **48**(3), 811–814 (2023).
14. J. Aubrecht, J. Pokorný, B. Švejkarová, M. Kamrádek, and P. Peterka, "Broadband thulium fiber amplifier for spectral region located beyond the L-band," Opt. Express **32**(10), 17932 (2024).
15. S. D. Agger and J. H. Povlsen, "Emission and absorption cross section of thulium doped silica fibers," Opt. Express **14**(1), 50 (2006).
16. P. Vařák, M. Leich, M. Kamrádek, J. Aubrecht, O. Podrazký, I. Bartoň, B. Švejkarová, A. Michalcová, K. Wondraczek, M. Jäger, I. Kašík, P. Peterka, and P. Honzátko, "Nanoparticle doping and molten-core methods towards highly thulium-doped silica fibers for 0.79 μm-pumped 2 μm fiber lasers – A fluorescence lifetime study," Journal of Luminescence **275**, 120835 (2024).